\documentclass[longbibliography,twocolumn,superscriptaddress]{revtex4-1}
\usepackage{amsmath,amssymb,graphicx,color,bm,siunitx} 

\begin{document}

\title{Optimal band gap for improved thermoelectric performance of\\
  two-dimensional Dirac materials}

\author{Eddwi~H.~Hasdeo}\email{eddw001@lipi.go.id}
\affiliation{Research Center for Physics, Indonesian Institute of
  Sciences (LIPI), Tangerang Selatan 15314, Indonesia}

\author{Lukas~P.~A.~Krisna}
\affiliation{Theoretical High Energy Physics and Instrumentation Research
  Group, Faculty of Mathematics and Natural Sciences, Institut
  Teknologi Bandung, Bandung 40132, Indonesia}

\author{Muhammad~Y.~Hanna}
\affiliation{Research Center for Physics, Indonesian Institute of
  Sciences (LIPI), Tangerang Selatan 15314, Indonesia}

\author{Bobby~E.~Gunara}
\affiliation{Theoretical High Energy Physics and Instrumentation Research
  Group, Faculty of Mathematics and Natural Sciences, Institut
  Teknologi Bandung, Bandung 40132, Indonesia}

\author{Nguyen~T.~Hung}
\affiliation{Frontier Research Institute of
  Interdisciplinary Sciences, Tohoku University, Sendai 980-8578,
  Japan}

\author{Ahmad.~R.~T.~Nugraha}\email{ahma079@lipi.go.id}
\affiliation{Research Center for Physics, Indonesian Institute of
  Sciences (LIPI), Tangerang Selatan 15314, Indonesia}
\affiliation{Department of Physics, Tohoku University,
  Sendai 980-8578, Japan}

\begin{abstract}
  Thermoelectric properties of two-dimensional (2D) Dirac materials
  are calculated within linearized Boltzmann transport theory and
  relaxation time approximation.  We find that the gapless 2D Dirac
  material exhibits poorer thermoelectric performance than the gapped
  one. This fact arises due to cancelation effect from electron-hole
  contributions to the transport quantities.  Opening the band gap
  lifts this cancellation effect.  Furthermore, there exists an
  optimal band gap for maximizing figure of merit ($ZT$) in the gapped
  2D Dirac material.  The optimal band gap ranges from $6k_B T$ to
  $18k_B T$, where $k_B$ is the Boltzmann constant and $T$ is the
  operating temperature in kelvin.  This result indicates the
  importance of having narrow gaps to achieve the best thermoelectrics
  in 2D systems.  Larger maximum $ZT$s can also be obtained by
  suppressing the lattice thermal conductivity.  In the most ideal
  case where the lattice thermal conductivity is very small, the
  maximum $ZT$ in the gapped 2D Dirac material can be many times $ZT$
  of commercial thermoelectric materials.
\end{abstract}

\date{\today}
\maketitle

\section{Introduction}

Thermoelectric (TE) materials convert temperature gradient into
electricity and thus they are useful for various devices utilizing
refrigerators and power
generators~\cite{goldsmid10-TEintro,heremans2013thermoelectrics}.
Unfortunately, it is well recognized that the efficiency of most of TE
materials is lower than other energy conversion systems so their
applications are still limited in the areas where the efficiency is
not an important issue.  To expand the applicability of
thermoelectrics, there has been extensive studies suggesting different
strategies with particular emphasis on improving the efficiency, such
as the energy band convergence~\cite{pei2011convergence}, the
hierarchical architecturing~\cite{biswas2012high}, and the
low-dimensional materials~\cite{hicks93-qweff,hung16-quantum}.
Theoretically, an efficient TE material should be a good electronic
conductor as well as a good thermal insulator.  The efficiency of
converting heat into electricity is related to the so-called TE figure
of merit, $ZT=S^2 \sigma T / \kappa$, where $S$, $\sigma$, $\kappa$
and $T$ are the Seebeck coefficient, electrical conductivity, thermal
conductivity, and operating temperature, respectively.  For many
decades, it has been a challenging issue even just to find materials
with $ZT \approx 1$ since $S$, $\sigma$ and $\kappa$ are generally
interrelated~\cite{majumdar04-TErev,vining09-TEinconvenient}.  In
other words, it is difficult to obtain a TE material with
simultaneously large $S$, large $\sigma$, and small $\kappa$ to
maximize $ZT$~\cite{mahan96-bestTE,zhou11-optbw,sofo94-optgap}.

Of different strategies to obtain better $ZT$, miniaturization of
materials has been an important route to enhance TE performance,
thanks to the quantum confinement effect that modifies the band
structure, effective mass, and density of
states~\cite{hicks93-qweff,hicks93-TE1D,hicks96-TEexp,hung16-quantum}.
Two-dimensional (2D) materials, in particular, are often suggested to
have better TE performance than bulk
materials~\cite{millie07-TEnewdir,zhang17-TE2D}.  A lot of research
efforts and grants have especially been invested on the 2D materials
whose electronic structure can be modeled by the Dirac
Hamiltonian~\cite{gomez16-2D}, or the so-called 2D Dirac materials.
The examples of 2D Dirac materials include, but not limited to,
graphene, silicene, germanene, transition metal dichalcogenides
(TMDs), and hexagonal boron nitride.  Some of them possess excellent
electronic and thermal properties, so one naturally hopes that the 2D
Dirac materials could be suitable for TE applications.  Some recent
findings may indicate the possibilities of 2D Dirac materials as a
good TE material.  For example, a large power factor, $S^2 \sigma$
(part of the numerator of $ZT$), has been either theoretically
proposed or experimentally observed in MoS$_2$ (one of
TMDs)~\cite{kedar17-MoS2TE} and in graphene-like
materials~\cite{yang14-silicene, duan2016high}.  Furthermore, by band
gap engineering, these 2D materials possess various values of the band
gaps, ranging from nearly zero to about
$2~\mathrm{eV}$~\cite{ni12-sige,zamin16-si,gomez16-2D}, so we can have
various choices of materials depending on the purpose.

However, despite the ``hype'' of current research in 2D Dirac
materials, their $ZT$s overall remains hard to push above unity.  In
particular, if we look at 2D semiconductors with moderate or wide band
gaps, the predicted or observed $ZT$ values are mostly less than
one~\cite{kedar17-MoS2TE,lee16-SnS2TE,wickramaratne15-inse}, while
those with smaller gaps tend to exhibit larger
$ZT$~\cite{ghaemi10,maassen13, zhao2014ultralow}.  This tendency
reminds us to some early works regarding the effect of band gap on the
$ZT$~\cite{mahan96-bestTE,sofo94-optgap}, in which it was suggested
that the best thermoelectrics in bulk (3D) systems can be obtained
with materials having narrow gaps~\cite{sofo94-optgap,gibbs2015band}.
Motivated by those works, we posit that there should also exist an
optimal gap (or a possible range of optimal gaps) for the 2D Dirac
materials in order to achieve the best thermoelectrics.  Once we know
the optimal gaps, the search for 2D materials with high TE
performance can be more focused on some certain materials.

In this paper, we will discuss TE properties of 2D Dirac materials
whose low energy dispersions are effectively described by the Dirac
Hamiltonian with band gap opening (closing) due to broken (unbroken)
inversion symmetry.  In these materials, electronic mobility is
generally larger for the smaller band gap.  As we will show later, the
excellent electron transport properties in these materials do not
automatically lead to efficient and high-performance thermoelectrics.
However, we can maximize the $ZT$ for the 2D Dirac materials by
considering the optimal gap.  From the calculations of the Seebeck
coefficient, electrical conductivity, thermal conductivity, and thus
$ZT$ within linearized Boltzmann transport theory and relaxation time
approximation (RTA), we find that the optimal gaps for the 2D Dirac
materials are about $6$--$18k_B T$.  Therefore, although recent trend
in TE research utilize moderate-gap or wide-gap 2D semiconductors as
potential TE
materials~\cite{kedar17-MoS2TE,lee16-SnS2TE,wickramaratne15-inse}, we
would suggest that it is better to use materials with narrow band gaps
within $6$--$18k_B T$ and then, if needed, we may further enhance
their TE performance by other techniques such as
doping~\cite{anno2017defect}, strain engineering~\cite{qin2017strain},
and manufacturing grain boundary or point defect~\cite{kim2015dense}
to diminish the phonon thermal conductivity.

\section{Model and methods}
\label{sec:th}

\begin{figure}[t!]
  \centering \includegraphics[clip,trim=0 9mm 0 0, width=8cm]{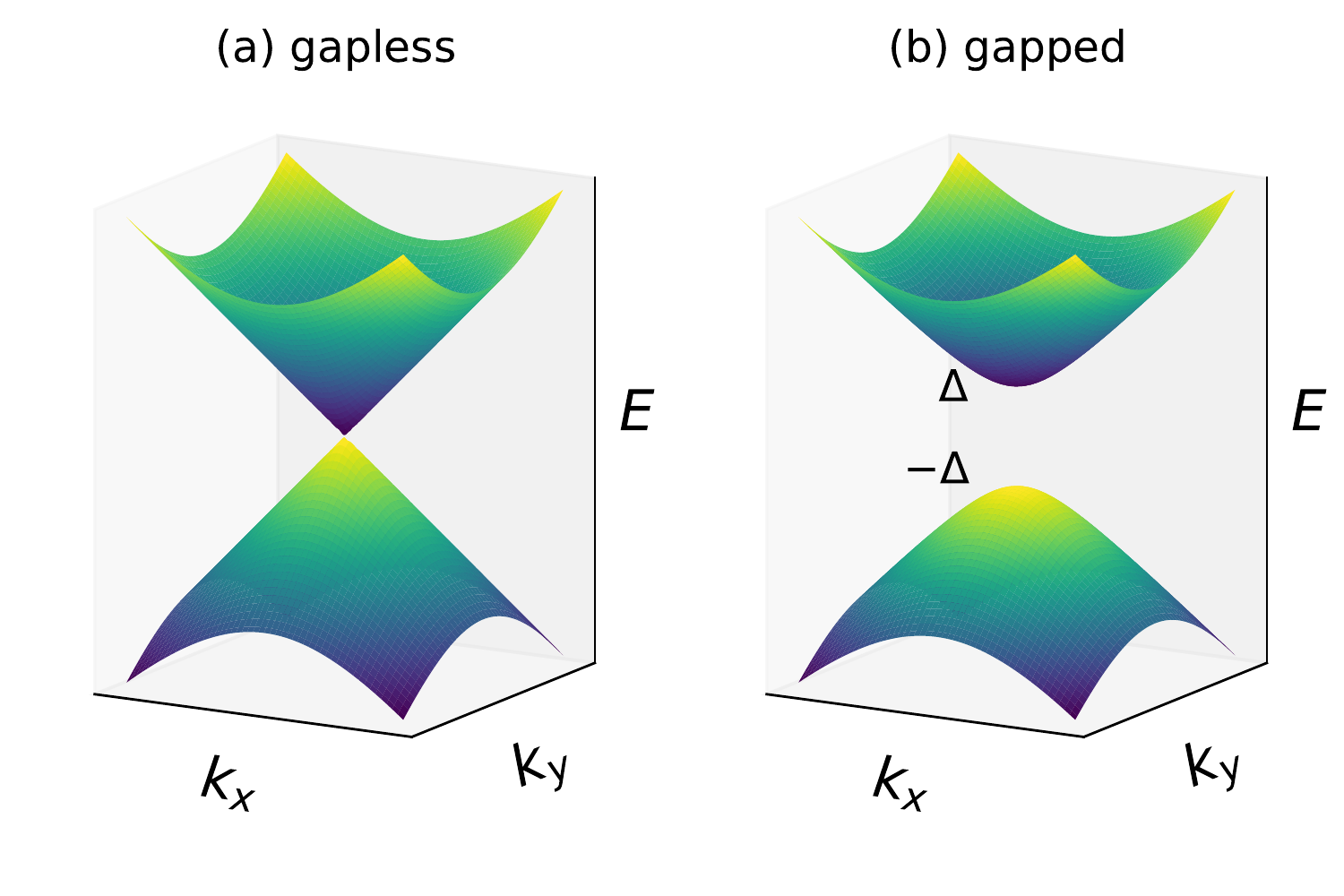}
  \caption{\label{figEdisp} Schematic two-band energy dispersion for
    (a) gapless and (b) gapped 2D Dirac materials.  Note that the band
    gap of the gapped Dirac material in this paper is expressed as
    $E_g = 2\Delta$.}
\end{figure}

For simplicity, we assume that the energy bands of gapless and gapped
Dirac materials are symmetric with respect to $E = 0$.  The 2D Dirac
material with a band gap $E_g = 2\Delta$ can be described by energy
dispersion,
\begin{equation}
  E(k) = \pm \sqrt{(\hbar v_F |\mathbf{k}|)^2 + \Delta^2},
  \label{eq:dirac}
\end{equation}
where $\hbar$ is the Planck constant, $v_F$ is the Fermi velocity, and
$\mathbf{k}$ is the 2D wave vector with magnitude
$k = |\mathbf{k}| = \sqrt{k_x^2 + k_y^2}$.  This energy dispersion is
illustrated in Figs.~\ref{figEdisp}(a) and (b) for gapless
($\Delta = 0$) and gapped (finite $\Delta$) 2D Dirac materials,
respectively.

We use the Boltzmann transport theory in the linear response regime
and apply the relaxation time approximation (RTA) for an isotropic
system, which is valid for our simplified model of 2D Dirac materials
whose transport properties do not depend on a particular orientation
in the 2D plane.  Within this approach, thermoelectric properties of a
2D Dirac material can be calculated from the transport coefficients
\begin{align}
  \label{eq:TEkernel}
  \mathcal{L}_i = \int_{-\infty}^\infty \mathcal{T} (E) (E-\mu)^i
  \left(-\frac{\partial f}{\partial E}\right) dE,
\end{align}
where $\mu$ is the chemical potential (or Fermi energy), $f(E)$ is the
Fermi-Dirac distribution, and $\mathcal{T}(E)$ is the transport
distribution function (TDF).  In Eq.~\eqref{eq:TEkernel}, $i$ takes a
value of $0$, $1$, or $2$ depending on the thermoelectric properties
to be calculated.

The explicit form of TDF is
\begin{align}
  \label{eq:tdfeq}
  \mathcal{T}(E)=v^2(E)\tau(E)\mathcal{D}(E),
\end{align}
where $v$ is the longitudinal velocity in a particular direction,
$\tau$ is the relaxation time, and $\mathcal {D}(E)$ is the density of
states (DOS).  For isotropic systems, $v$ can be related with group
velocity $v_g$ and dimension $d$ according to the relation
$v^2 = v_g^2 / d$, where $v_g = \hbar^{-1}(dE/dk)$. Therefore, we can
express $v$ for the 2D Dirac materials ($d = 2$) as:
\begin{align}
  v^2 = \frac{v_F^2}{2}\left(\frac{E^2 - \Delta^2}{E^2}\right).
\end{align}

For the energy-dependent relaxation time, we assume that short-range
impurity scattering dominates the relaxation mechanism and as result
the relaxation time is inversely proportional to the
DOS~\cite{hung17inse,hung2018universal}, i.e.,
\begin{align}
  \label{eq:trelax}\tau(E) = C \left[\mathcal{D}(E)\right]^{-1},
\end{align}
where $C$ is the scattering coefficient (in units of
$\mathrm{W}^{-1}\mathrm{m}^{-3}$) that depends on the material
dimension and confinement length~\cite{hung2018universal}.  This
assumption can be derived from Fermi's golden rule and is suitable for
the scattering mechanism involving electron-phonon interactions where
either acoustic or optical phonons scattered by electrons within
temperature range
$300$--$700$~K~\cite{hung2018universal,hung17inse,zhou2011optimal}.
For the sake of completeness, in Appendix~\ref{app:crta} we also
present the calculation result within a constant $\tau_0$ (relaxation
time independent of $E$)~\cite{githubdiracTE}, which plays a
complementary role to $\tau(E)$.  Note that, under the assumption of
energy-dependent relaxation time [Eq.~\eqref{eq:trelax}], the DOS will
disappear from the TDF of Eq.~\eqref{eq:tdfeq}.  However, under the
constant relaxation time approximation (CRTA) the DOS term will remain
and is given in Eq.~\eqref{eq:dosdirac2d}.

Using the transport coefficients, one can calculate the electrical
conductivity $\sigma$, Seebeck coefficient $S$, and electron thermal
conductivity $\kappa_e$.  They are respectively given by
\begin{align}
  \label{eq:defsigma}
  \sigma &= q^2\mathcal{L}_0,\\
  S &= \frac{1}{qT} \frac{\mathcal{L}_1}{\mathcal{L}_0},
\end{align}
and
\begin{align}
  \label{eq:defkappa}
  \kappa_e &= \frac{1}{T}\left(\mathcal{L}_2-
             \frac{\left(\mathcal{L}_1\right)^2}{\mathcal{L}_0}\right),
\end{align}
From these thermoelectric properties, along with the lattice thermal
conductivity $\kappa_{\mathrm{ph}}$, we can calculate the
thermoelectric figure of merit,
\begin{align}
  ZT &= \frac{S^2 \sigma}{\kappa_e + \kappa_{\mathrm{ph}}} T.
\end{align}
Whenever necessary, especially to simplify some equations, we will use
reduced (dimensionless) variables for the energy dispersion
$  \varepsilon = E/k_B T$,
and also for the chemical potential, $ \eta = \mu/k_B T$.

The transport coefficients $\mathcal{L}_i$ are generally calculated by
considering all bands available within $E = [-\infty,\infty]$.
However, in most of materials, thermoelectric properties are dominated
by the states near the Fermi energy.  In this work, we adopt the
two-band model involving a valence band and a conduction band as a
minimum requirement to include the bipolar effect (or sign inversion
of the Seebeck coefficient), which is always observed in materials
having two different types of carriers, i.e., electrons and holes in
the conduction and valence bands, respectively.  A justification for
the two-band model is given in Appendix~\ref{app:dft} by comparing it
with the multiband, first-principles calculation.  In the two-band
model, $\sigma$, $S$, and $\kappa_e$ can be written as
follows~\cite{goldsmid10-TEintro}:
\begin{align}
  \label{eq:2bandsig}
  \sigma &= \sigma_c + \sigma_v,\\
  \label{eq:2bandS}
  S &=
      \frac{\sigma_c S_c + \sigma_v S_v}{\sigma_c + \sigma_v},\\
  \label{eq:2bandkap}
  \kappa_e &= \frac{\sigma_c \sigma_v}{\sigma_c + \sigma_v}(S_c - S_v)^2
             + (\kappa_{e,c} + \kappa_{e,v}),
\end{align}
where the additional subscript $c$ ($v$) labels the conduction
(valence) band.

Based on Eqs.~\eqref{eq:2bandsig}--\eqref{eq:2bandkap}, the integral
$\mathcal{L}_i$ in Eqs.~\eqref{eq:defsigma}--\eqref{eq:defkappa} can
be decomposed into $\mathcal{L}_{i,c}$ and $\mathcal{L}_{i,v}$ with
the integration over energy intervals $[0,\infty]$ and $[-\infty,0]$
for the conduction and valence bands, respectively, i.e.,
\begin{align}
  \label{eq:int-cond}
  \mathcal{L}_{i,c} = \int_{0}^\infty \mathcal{T} (E) (E-\mu)^i
  \left(-\frac{\partial f}{\partial E}\right) dE .
\end{align}
and
\begin{align}
  \label{eq:int-val}
  \mathcal{L}_{i,v} = \int_{-\infty}^0 \mathcal{T} (E) (E-\mu)^i
  \left(-\frac{\partial f}{\partial E}\right) dE .
\end{align}

\section{Thermoelectrics of 2D Dirac materials}
\label{sec:gapped}

The transport distribution function of 2D Dirac materials
is given by
\begin{align}
  \label{eq:tdfdirac}
\mathcal{T}(E) &= \frac{C v_F^2}{2} \left(\frac{E^2 - \Delta^2}{E^2} \right),
\end{align}
with the energy intervals $[-\infty, -\Delta]$ and $[\Delta, \infty]$
to be considered in the calculation of the TE integrals
Eqs.~\eqref{eq:int-cond} and \eqref{eq:int-val}.  We also define the
dimensionless gap, $\widetilde{\Delta} = \Delta/k_B T$.  The TE
integral for the conduction band is then expressed as
\begin{align}
  \mathcal{L}_{i,c} 
  = \frac{Cv_F^2(k_BT)^i}{2} \left(
  \mathcal{F}_{i,c}(\eta-\widetilde{\Delta})-
  \mathcal{G}_{i,c}(\widetilde{\Delta},\eta)\right),
\end{align}
with
\begin{align}
  \mathcal{F}_{i,c} (\eta)
  &= \int_{-\eta}^\infty
    \frac{x^i e^{x}}{\left(e^{x} +1 \right)^2} dx, \label{eq:fic}\\
    \mathcal{G}_{i,c}(\widetilde{\Delta},\eta)
  &= \int_{\widetilde\Delta-\eta}^\infty
    \frac{\widetilde\Delta^2}{(x+\eta)^2} \frac{x^i e^x}{(e^x+1)^2}
    dx\label{eq:gic}.
\end{align}
Similarly, the TE integral for the valence band is
\begin{align}
  \mathcal{L}_{i,v} = \frac{Cv_F^2(k_BT)^i}{2} \left(
  \mathcal{F}_{i,v}(\eta+\widetilde{\Delta})
  -\mathcal{G}_{i,v}(\widetilde{\Delta},\eta)\right),
\end{align}
with
\begin{align}
  \mathcal{F}_{i,v} (\eta)
  &= \int_{-\infty}^{-\eta}
    \frac{x^i e^{x}}{\left(e^{x}
    +1 \right)^2} dx,\label{eq:fiv}\\
  \mathcal{G}_{i,v}(\widetilde{\Delta},\eta)
  &= \int_{-\infty}^{-\widetilde{\Delta}-\eta}
    \frac{\widetilde\Delta^2}{(x+\eta)^2}
    \frac{x^i e^x}{(e^x+1)^2} dx.\label{eq:giv}
\end{align}

The $\mathcal{F}_i$ integrals Eqs.~\eqref{eq:fic} and \eqref{eq:fiv}
can be obtained analytically.  They are:
\begin{align}
  \mathcal{F}_{0,c} (\eta) =& \frac{e^\eta}{e^\eta + 1},\\
  \mathcal{F}_{1,c} (\eta) =& \frac{\eta}{e^\eta + 1} + \ln(1 + e^{-\eta}),\\
  \mathcal{F}_{2,c} (\eta) =& \frac{\pi^2}{3} 
                             - \frac{\eta^2}{e^\eta + 1}
                             - 2\eta \ln(1 + e^{-\eta})\notag\\
                           & + 2 \mathrm{Li}_2(-e^{-\eta}),\\
  \mathcal{F}_{0,v} (\eta) =& \frac{1}{e^\eta + 1},\\
  \mathcal{F}_{1,v} (\eta) =& - \frac{\eta}{e^\eta + 1} - \ln(1 + e^{-\eta}),\\
  \mathcal{F}_{2,v} (\eta) =& \frac{\eta^2}{e^\eta + 1}
                             + 2x \ln(1 + e^{-\eta})
                             - 2 \mathrm{Li}_2(-e^{-\eta}),
\end{align}
where
$\mathrm{Li}_j(z) = \displaystyle\sum_{n=1}^\infty \frac{z^n}{n^j}$ is
the polylogarithmic function. Note that $\mathcal{F}_{i,c}$ and
$\mathcal{F}_{i,v}$ are connected via electron-hole symmetry of the
system:
\begin{subequations}
\label{eq:eh-sym}
\begin{align}
  \mathcal{F}_{0,c} (\eta) &= \mathcal{F}_{0,v} (-\eta)\\
  \mathcal{F}_{1,c} (\eta) &= -\mathcal{F}_{1,v} (\eta)\\
  \mathcal{F}_{2,c} (\eta) &= \mathcal{F}_{2,v} (-\eta)
\end{align}
\end{subequations}
Unlike $\mathcal{F}_i$, the $\mathcal{G}_i$ integrals for the
conduction and valence band contributions cannot be calculated
analytically, thus we employ numerical integrations to obtain
$\mathcal{G}_i$~\cite{githubdiracTE}.

\subsection{Gapless case}
Let us firstly discuss the gapless case.  In the gapless limit,
$\widetilde\Delta=0$ so that $\mathcal{G}_i$ is also equal to zero.
As a result, electronic contributions from the conduction band of
gapless Dirac materials to the TE quantities are solely a function of
$\mathcal{F}_i$:
\begin{align}
  \sigma_c &= q^2 \mathcal{L}_{0,c} \notag\\
           &= \frac{1}{2} e^2 v_F^2 C \mathcal{F}_{0,c} (\eta),\\
  S_c &= \frac{1}{qT} \frac{\mathcal{L}_{1,c}}{\mathcal{L}_{0,c}} \notag\\
           &= - \frac{k_B}{e} \frac{\mathcal{F}_{1,c}(\eta)}{
             \mathcal{F}_{0,c}(\eta)},\\
  \kappa_{e,c} &= \frac{1}{T} \left(\mathcal{L}_{2,c} -
                 \frac{(\mathcal{L}_{1,c})^2}{\mathcal{L}_{0,c}}\right)
                 \notag\\
           &= \frac{1}{2} v_F^2 C k_B^2 T \left(\mathcal{F}_{2,c} (\eta) -
             \frac{(\mathcal{F}_{1,c}(\eta))^2}{\mathcal{F}_{0,c}(\eta)} \right),
\end{align}
where $e \approx 1.602 \times 10^{-19}~\mathrm{C}$ is the elementary
charge.  For convenience, hereafter we will use the units
$S_0 = k_B / e$ ($\sim$$87$ \si{\micro}V/K), $\sigma_0 = e^2 v_F^2 C /
2$, and $\kappa_0 = v_F^2 C k_B^2 T / 2$.  Note that
$C$ depends on the confinement length or thickness of 2D
materials~\cite{hung2018universal}. Therefore, the natural units of
electrical conductivity and thermal conductivity, i.e.,
$\sigma_0$ and $\kappa_0$, are thickness-dependent.

The contribution of the valence band to the thermoelectric
quantities $\sigma_v$, $S_v$, and
$\kappa_{e,v}$, can be expressed similarly with that of the conduction
band:
\begin{align}
  \sigma_v &= \sigma_0 \mathcal{F}_{0,v} (\eta),\\
  S_v &= - S_0 \frac{\mathcal{F}_{1,v}(\eta)}{\mathcal{F}_{0,v}(\eta)},\\
  \kappa_{e,v} &= \kappa_0 \left(\mathcal{F}_{2,v} (\eta)
                 - \frac{(\mathcal{F}_{1,v}(\eta))^2}{
                 \mathcal{F}_{0,v}(\eta)} \right).
\end{align}
 We  set the lattice thermal conductivity as an adjustable
quantity,
\begin{align}
  \kappa_{\mathrm{ph}} = r_\kappa \kappa_0,
\end{align}
where
$r_\kappa$ is a material parameter and may be engineered to maximize
the $ZT$.  We assume that $T$ and $\eta$ dependences of $\kappa_{\rm
  ph}$ are fully adjusted to the free parameter $r_\kappa$.

\begin{figure}[t!]
  \centering \includegraphics[clip,width=8cm]{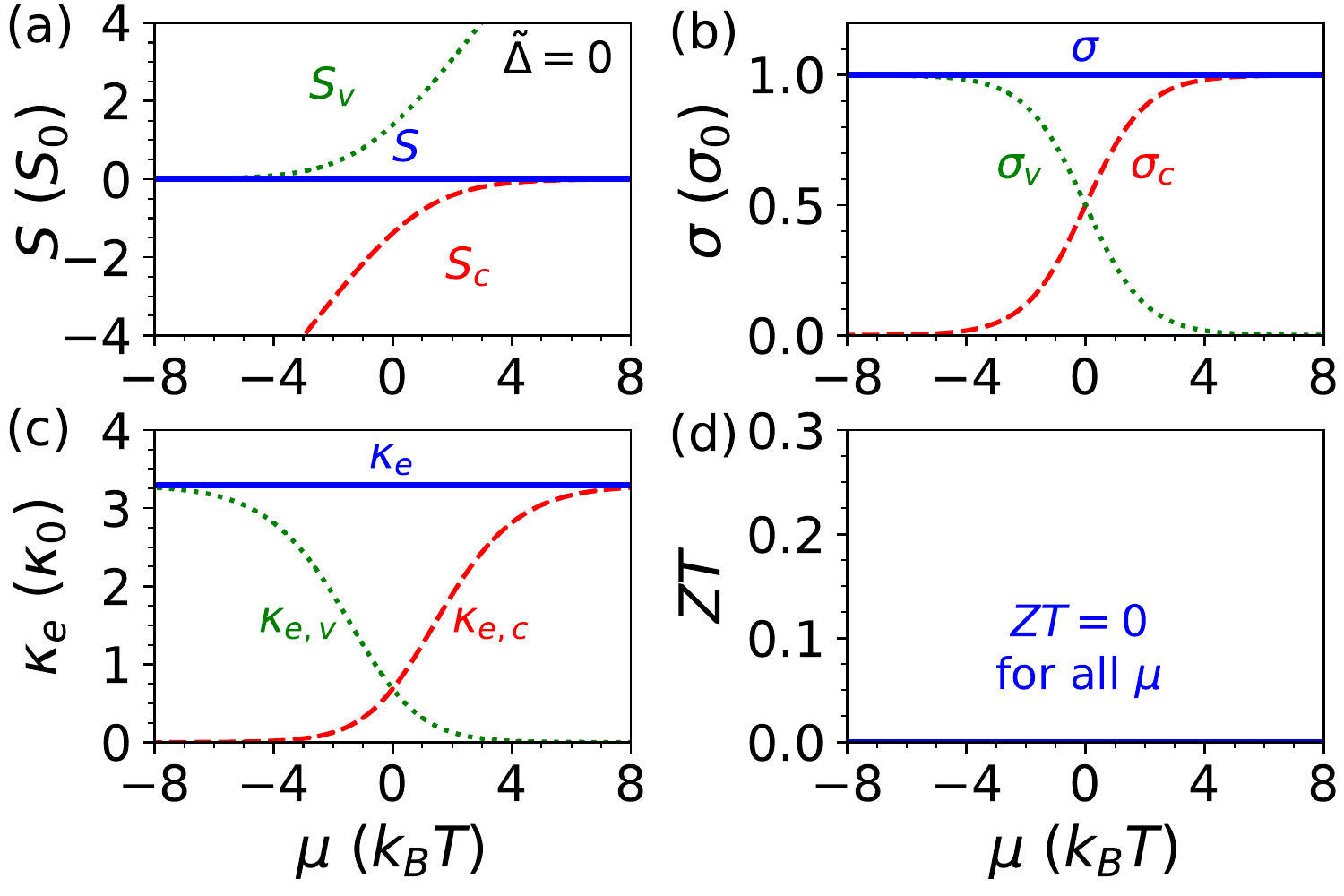}
  \caption{\label{figTEgapless} Thermoelectric properties of gapless
    2D Dirac material as a function of chemical potential.  (a)
    Seebeck coefficient in units of $S_0 = k_B / e$.  (b) Electrical
    conductivity in units of $\sigma_0 = e^2 v_F^2 C/2$.  (c) Electron
    thermal conductivity in units of $\kappa_0 = v_F^2 C k_B^2 T$.
    (d) Dimensionless figure of merit.}
\end{figure}

Based on the formulas of the two-band model
[Eqs.~\eqref{eq:2bandsig}--\eqref{eq:2bandkap}] and the electron-hole
symmetry in the TE integrals [Eq.~\eqref{eq:eh-sym}], it is obvious
that the Seebeck coefficient $S$ becomes zero [see
  Fig.~\ref{figTEgapless}(a)], whereas the electrical conductivity
$\sigma$ and electron thermal conductivity $\kappa_e$ have constant
values for whole $\eta$ [see Figs.~\ref{figTEgapless}(b) and (c)].  As
a result, for the gapless 2D Dirac material considered in this
approximation, we have $ZT(\eta) = 0$ .

The dotted and dashed lines in Figs.~\ref{figTEgapless}(a)--(c)
display the sole contributions of valence and conduction bands to $S$,
$\sigma$, and $\kappa_e$, respectively. The dramatic cancelation of
the total Seebeck coefficient arising from the electron-hole symmetry
in Eqs.~\eqref{eq:eh-sym}.  Since the relaxation time is assumed to be
inversely proportional to DOS in Eq.~\eqref{eq:trelax}, the TDF
becomes energy-independent and the Seebeck coefficient completely
vanishes for all doping levels.  We note that the energy-dependent
$\tau(E)$ [Eq.~\eqref{eq:trelax}] fails to describe the region near
the Dirac point ($\mu=0$) as the DOS vanishes.  In the realistic
situation $\tau(E)$ should be finite even at $\mu=0$ so that the
accidental, perfect cancellation should not appear.

We clarify that the CRTA, which is presented in
Appendix~\ref{app:crta}, gives nonzero and finite value of $S$ for the
gapless 2D Dirac material, consistent with an earlier work by Sharapov
and Varlamov~\cite{sharapov12}.  It is expected that constant
relaxation time $\tau_0$ and energy-dependent relaxation time
$\tau(E)$ play a complementary role giving the total lifetime
$\tau_{\rm tot}^{-1}(E)=\tau^{-1}(E)+\tau_{0}^{-1}$.  Nevertheless,
with the sole use of $\tau(E)$, we expose the demerit of electron-hole
cancellation to the TE properties.  To get rid of the poor TE
performance due to the electron-hole cancellation, we may apply a
magnetic field as a facilitator to break the electron-hole symmetry
and thus obtain a larger, nonsaturating Seebeck
coefficient~\cite{skinner2018large}.  However, applying the magnetic
field on the order of several teslas is beyond the current practical
capability of the TE industry.  A simpler way to lift this
electron-hole cancellation effect is by using \emph{gapped} 2D Dirac
materials.

\begin{figure}[t!]
  \centering \includegraphics[clip,width=8cm]{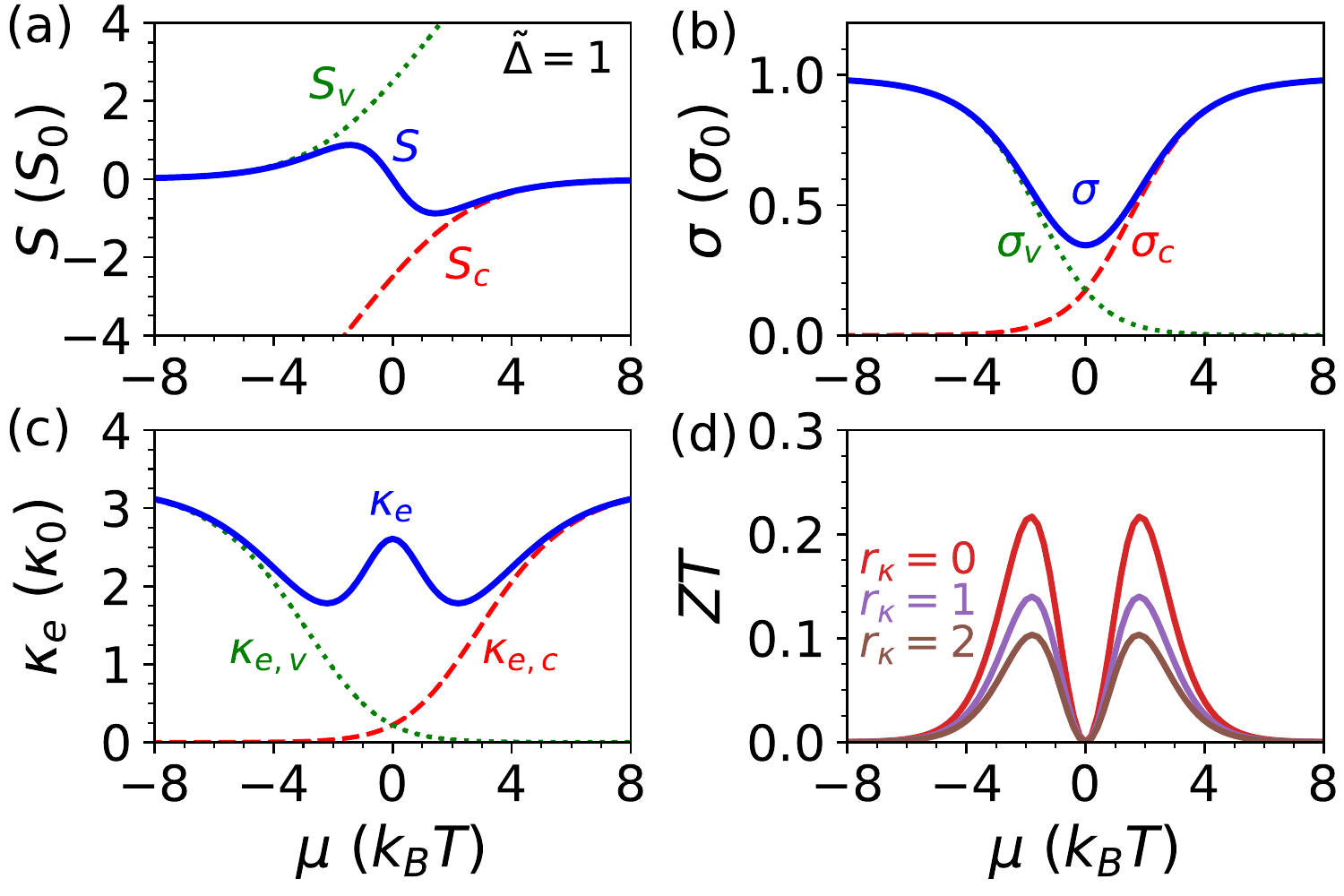}
  \caption{\label{figTEgapped} Thermoelectric properties of a 2D Dirac
    material with $\widetilde{\Delta} = 1$ (equivalently, band gap $E_g =
    2 k_B T$).  (a) Seebeck coefficient in units of $S_0 = k_B / e$.
    (b) Electrical conductivity in units of $\sigma_0 = e^2 v_F^2
    C/2$.  (c) Electron thermal conductivity in units of $\kappa_0 =
    v_F^2 C k_B^2 T$.  (d) Dimensionless figure of merit calculated
    with three different parameters $r_\kappa$ (representing lattice
    thermal conductivity).}
\end{figure}

\subsection{Gapped case}

Figure~\ref{figTEgapped} depicts the TE properties of the 2D gapped
Dirac material for $\Delta=k_BT$. The calculation result of Seebeck
coefficient shows finite values whose sign changes across the charge
neutrality point, indicating competing contribution of electron and
hole, as represented by dotted and dashed lines in
Fig.~\ref{figTEgapped}(a).  Nevertheless, the perfect cancelation has
now been avoided, thanks to the gap opening.  The electrical
conductivity exhibits a dip in the band gap and saturates to a finite
value far away from the gap [Fig.~\ref{figTEgapped}(b)].  The electron
thermal conductivity has a local maximum in the small gap limit due to
the combination of of $S_{c,v}$ and $\sigma_{c,v}$ that add each
other.  The resulting $ZT$ peaks appear slightly above the band edges
with a maximum value of about $0.2$ for $\widetilde\Delta=1$ if no
phonon contributes to the thermal conductivity ($r_\kappa=0$).  The
$ZT$ peaks monotonically decrease as the phonon contribution to the
thermal conductivity increases.

\begin{figure}[t]
  \centering \includegraphics[clip,width=8cm]{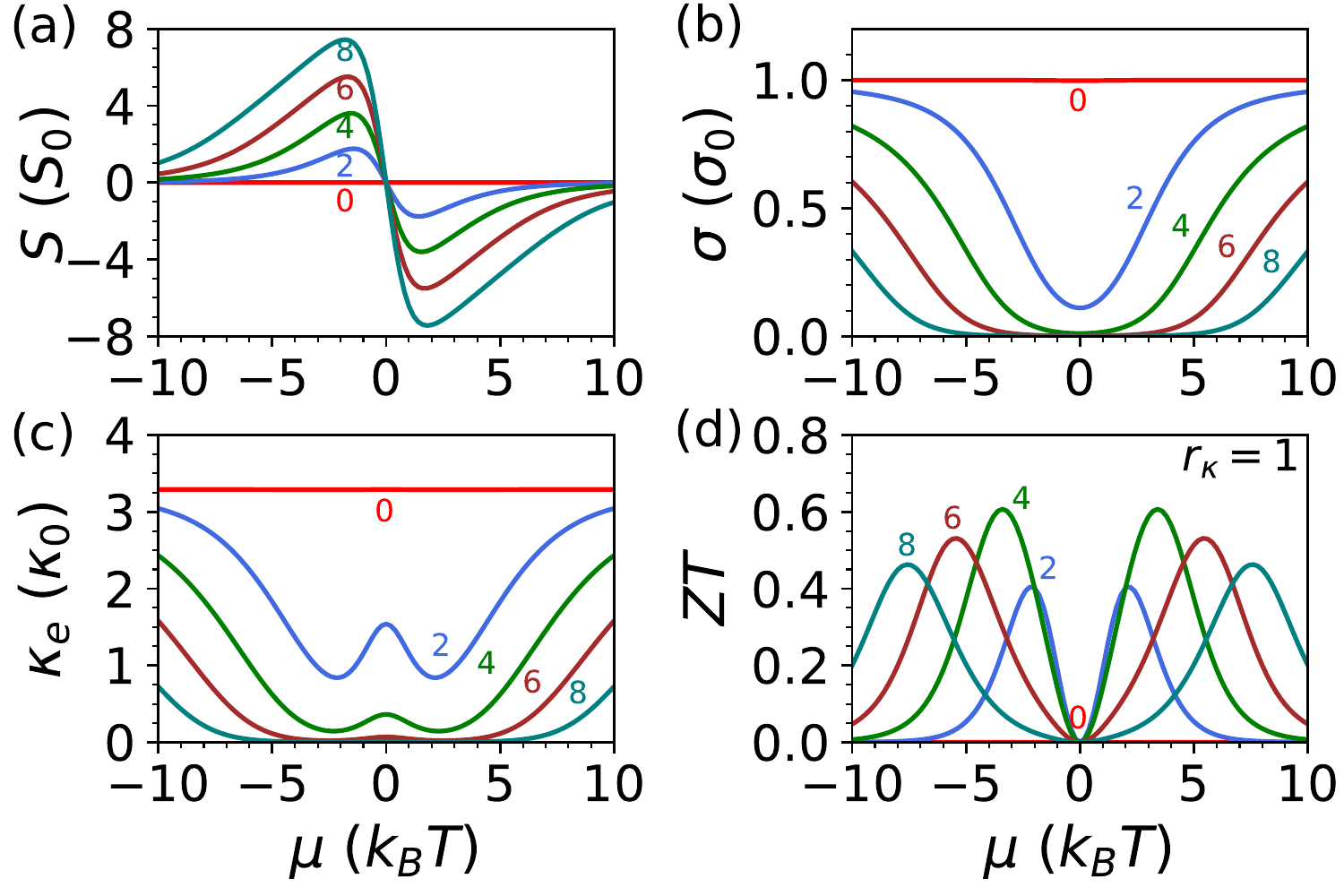}
  \caption{\label{figTEvarygap} Thermoelectric properties of 2D Dirac
    materials calculated for several different gaps, $\widetilde{\Delta} =
    0, 2, 4, 6, 8$.  (a) Seebeck coefficient in units of $S_0 = k_B
    / e$.  (b) Electrical conductivity in units of $\sigma_0 = e^2
    v_F^2 C/2$.  (c) Electron thermal conductivity in units of
    $\kappa_0 = v_F^2 C k_B^2 T$.  (d) Dimensionless figure of merit
    $ZT$ with $r_\kappa = 1$.  Note that the numbered curves in each
    figure are used to distinguish $\widetilde{\Delta} = 0, 2, 4, 6, 8$,
    which are equivalent to $E_g / k_B T = 0, 4, 8, 12, 16$.}
\end{figure}

\section{Optimal band gap for better thermoelectrics}
\label{sec:optimal}

Understanding how band gap alters the TE properties is of importance
for designers of TE materials.  To see the evolution of TE properties
with respect to the band gap, we show in Fig.~\ref{figTEvarygap} the
calculation results of $S$, $\sigma$, $\kappa_e$, and $ZT$ for
$\widetilde\Delta=0$--$8$.  As the band gap increases, the Seebeck
coefficient monotonically increases while electrical and electron
thermal conductivities decrease.  Combined action of these
interrelated quantities shall give $ZT$ that can be maximized by
tuning $\widetilde\Delta$.  For $r_\kappa=1$, the $ZT$ value reaches
maximum near the band edge and becomes largest at $\widetilde\Delta=4$
or corresponding band gap $E_g=8k_BT$ [Fig.~\ref{figTEvarygap}(d)].

\begin{figure}[t!]
  \centering \includegraphics[clip,width=8cm]{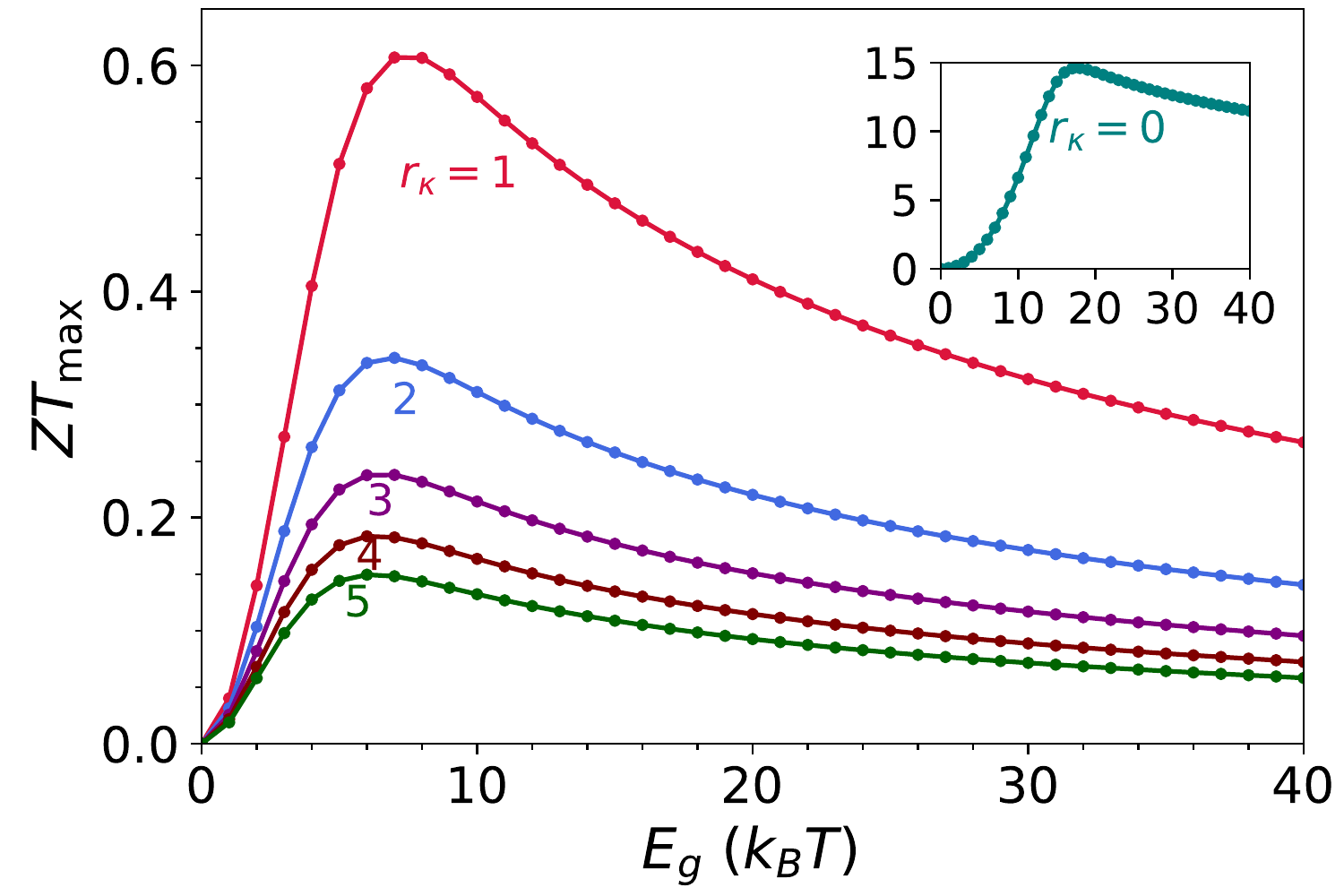}
  \caption{\label{figZTopt} Maximum $ZT$ values versus band gaps of 2D
    Dirac materials for several different lattice thermal conductivity
    parameters $r_\kappa$.  Inset shows the result for $r_\kappa = 0$
    which might be unrealistic but it gives the largest possible
    $ZT$.}
\end{figure}

To find the optimal band gap for the 2D Dirac materials, we can
numerically calculate the maximum $ZT$ value scanned through various
doping levels $\mu$ and then plot $ZT_{\textrm{max}}$ (the maximum
$ZT$ found after scanning $\mu$) as a function of $E_g$ and
$r_\kappa$.  The result is shown in Fig.~\ref{figZTopt}.  We can see
that $ZT_{\textrm{max}}$ typically peaks at $E_g=6$--$18k_BT$
depending on phonon thermal conductivity coefficient $r_\kappa$.  By
the increase of $r_\kappa$, we find that $ZT_{\textrm{max}}$ tends to
be achieved at smaller band gap, saturated at $E_g \sim 6k_B T$.
Decreasing phonon thermal conductivity through different methods such
as defect engineering, heterostructures and strain is favorable to
enhance $ZT$.  For example, the ultralow thermal conductivity denoted
by $r_\kappa = 0$ will give the largest possible $ZT_{\textrm{max}}$
far above unity and it favors larger band gap of about $18k_BT$ as
shown in the inset of Fig.~\ref{figZTopt}.

It should be noted that the $ZT_{\textrm{max}}$ profile as a function
of $E_g$ is sensitive to $\tau(E)$ function and is also affected by
the band structure.  The $ZT_{\textrm{max}}$ profile for a given
$r_\kappa$ in the case of 2D Dirac materials has a peak when assuming
$\tau(E)\propto{\rm DOS}^{-1}$ due to the deterioration of bipolar
effect, which means that the valence and conduction bands do not mix
in the TE transport.  A power law decay with the exponential factor of
$1/\Delta$ for $ZT_{\textrm{max}}$ at large band gaps is a hallmark of
$\tau(E)\propto{\rm DOS}^{-1}$ approximation for 2D gapped Dirac
materials.  This feature follows from asymptotic behavior of
$\mathcal{G}_{i,c}\rightarrow 1/\widetilde{\Delta}$ in
Eq.~\eqref{eq:gic} when the chemical potential reach the band edge
($\eta=\widetilde\Delta$).  In Appendixes~\ref{app:crta}
and~\ref{app:para}, we show different $ZT_{\textrm{max}}$ profiles for
2D Dirac and parabolic-band materials within the CRTA that do not show
clear peaks of $ZT_\textrm{max}$.  However, we find that the optical
chemical potential related to the $ZT_{\textrm{max}}$ is always very
close to the band edge when $E_g > 5 k_BT$ and then the starting gaps
to enhance $ZT_\textrm{max}$ appear around $10k_BT$, which is still
within $6$--$18k_BT$.  Further works are desired to check whether the
optimal gaps will appear beyond the range we find in the present
study.

\section{Conclusions and Perspective}
\label{sec:con}
We have shown that optimal band gaps within $6$--$18k_BT$ are useful
for maximizing $ZT$ in the 2D Dirac materials, where the TE properties
were calculated considering the energy-dependent relaxation time
$\tau (E)$ inversely proportional to the DOS.  This conclusion is
supplemented by considering the constant relaxation time approximation
in Appendixes, as well as comparison with the parabolic-band
materials.  Our calculations indicate that the gapless 2D Dirac
material is not good for thermoelectrics, but opening a gap of few
$k_BT$ is beneficial to enhance its TE properties.

The candidates of the gapped 2D Dirac materials that are suitable for
TE applications may already possible to fabricate, such as bilayer
graphene (with electrically tunable gap) and commensurate graphene-hBN
heterostructure~\cite{hunt2013massive,duan2016high} whose minigap
about tens of meV emerges due to inversion symmetry breaking.  In
particular, the graphene-hBN heterostructure has excellent
$S^2\sigma$~\cite{duan2016high}.

Our model is also readily extensible for other systems by using
appropriate parameters, so that it may trigger further theoretical
works on thermoelectrics. Multiband effect which is not considered in
this calculation might enhance $ZT$ and can be incorporated in a
straight-forward fashion within linearized Boltzmann transport
equation.

\begin{acknowledgements}
  E.H. acknowledges research grant from Indonesia Toray Science
  Foundation.  L.P.A., and B.E.G. acknowledge the financial support
  from Ministry of Research, Technology, and Higher Education of the
  Republic of Indonesia (Kemenristekdikti) through PDUPT 2018--2019. 
  N.T.H. acknowledge the financial support from the Frontier Research 
  Institute for Interdisciplinary Sciences, Tohoku University.
  N.T.H. and A.R.T.N are grateful to Prof. R. Saito (Tohoku
  University) for the fruitful discussion on thermoelectrics in
  several recent years.
\end{acknowledgements}

\appendix

\section{Results within constant relaxation
  time approximation (CRTA)}
\label{app:crta}

Here we present the calculation of TE properties of 2D Dirac materials
within the constant relaxation time approximation (CRTA).  The results
within this approximation overestimate the value of $S$ and $\sigma$
in comparison to those given in the main text.  As a result, the
gapless 2D Dirac material will have finite (nonzero) $S$, unlike what
we have shown in the main text.

In the CRTA, the transport distribution function for the 2D Dirac
materials is given by
\begin{equation}
  \mathcal{T}(E)= \tau_0 \left[\frac{v_F^2}{2}
    \left(\frac{E^2-\Delta^2}{E^2}\right)\right] \mathcal{D}(E),
\end{equation}
where $\tau_0$ is the relaxation time constant. The density of states
is defined by
\begin{align}
  \label{eq:dosdirac2d}
  \mathcal{D}(E) =\frac{g|E|}{2\pi L (\hbar v_F)^2}\Theta(|E|-|\Delta|),
\end{align}
where $g$ is degeneracies, $L$ is the confinement length, and
$\Theta(x)$ is the Heaviside step function, i.e., $\Theta(x)=1$ if
$x>0$ and $\Theta(x)=0$ otherwise.

\begin{figure}[t]
  \centering \includegraphics[clip,width=8cm]{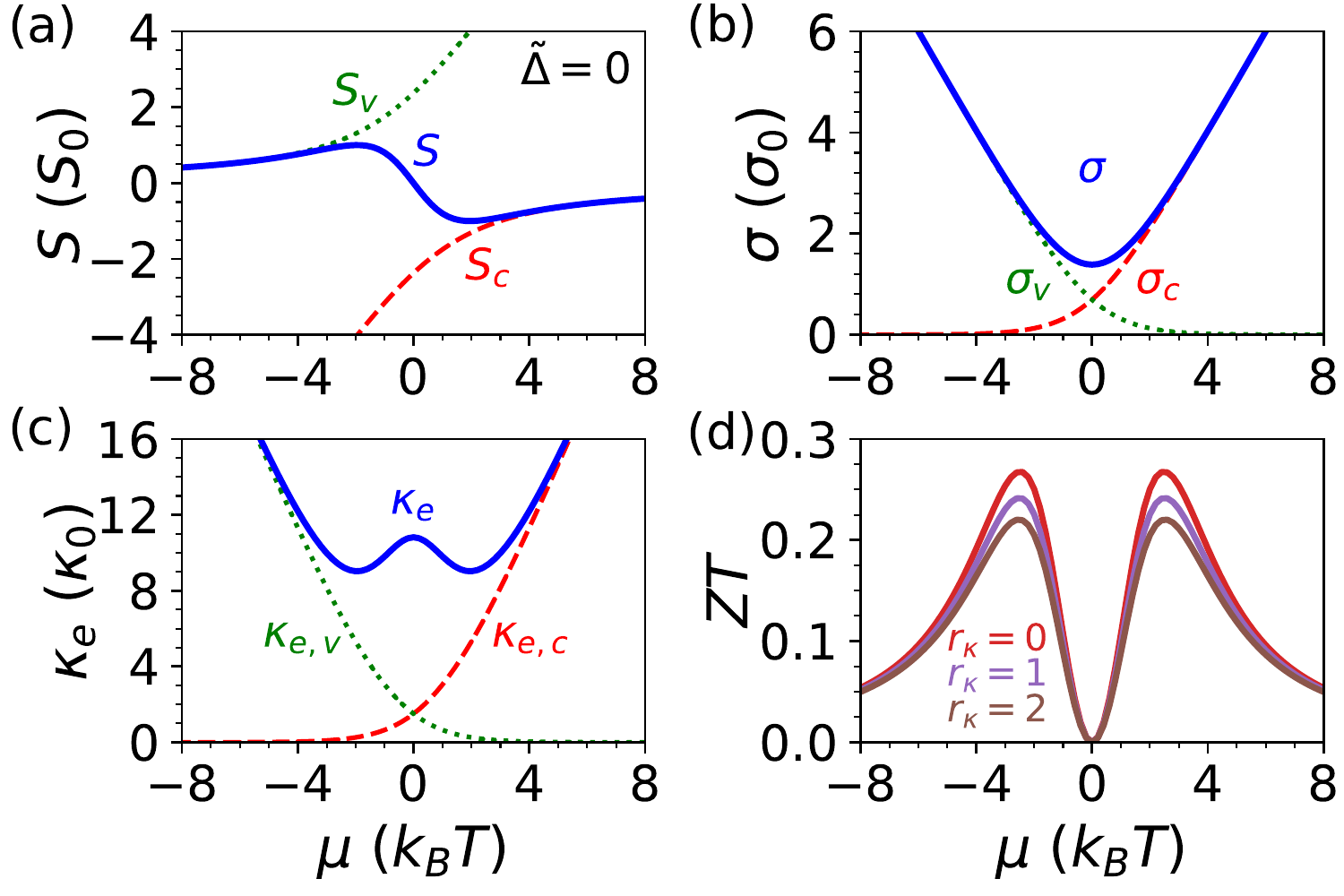}
  \caption{\label{CRTfigTEgapless} Thermoelectric properties of
    gapless 2D Dirac material as a function of chemical potential
    within the CRTA.  (a) Seebeck coefficient in units of
    $S_0 = k_B / e$.  (b) Electrical conductivity in units of
    $\sigma_0 = g \tau_0 e^2 k_B T /(4\pi \hbar^2 L)$.  (c) Electron
    thermal conductivity in units of
    $\kappa_0 = g \tau_0 k_B^3 T^2/(4\pi \hbar^2 L)$.  (d)
    Dimensionless figure of merit.}
\end{figure}

\begin{figure}[t]
  \centering \includegraphics[clip,width=8cm]{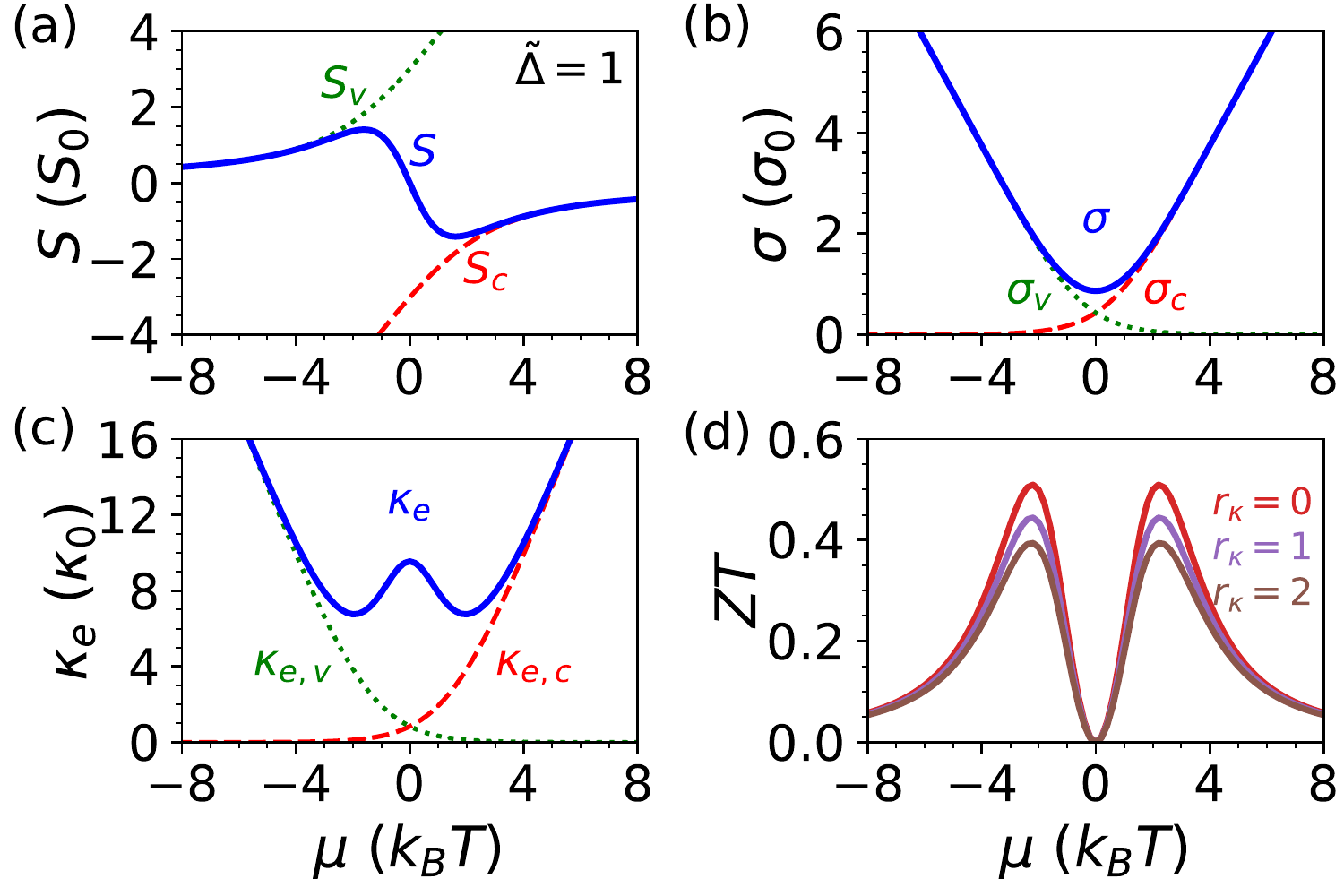}
  \caption{\label{CRTfigTEgapped} Thermoelectric properties of a 2D
    Dirac material within the CRTA for $\widetilde{\Delta} = 1$
    (equivalently, band gap $E_g = 2 k_B T$).  (a) Seebeck coefficient
    in units of $S_0 = k_B / e$.  (b) Electrical conductivity in units
    of $\sigma_0 = g \tau_0 e^2 k_B T/(4\pi\hbar^2 L)$.  (c) Electron
    thermal conductivity in units of
    $\kappa_0 = g \tau_0 k_B^3 T^2/(4\pi\hbar^2 L)$.  (d)
    Dimensionless figure of merit calculated with three different
    parameters $r_\kappa$ (representing lattice thermal
    conductivity).}
\end{figure}

After some algebra~\cite{githubdiracTE}, we can obtain the TE integral
for the conduction band of the 2D Dirac material within the CRTA as
\begin{align}
  \label{eq:crtacon}
  \mathcal{L}_{i,c} =&\frac{g \tau_0}{4\pi\hbar^2 L}(k_BT)^{i+1}
                       \Big[\mathcal{F}_{i+1,c}(\eta-
                       \widetilde{\Delta})\nonumber\\
                    &+\eta\mathcal{F}_{i,c}(\eta-\widetilde{\Delta})
                      -\widetilde{\mathcal{G}}_{i,c}
                      (\widetilde{\Delta},\eta)\Big].
\end{align}
Similarly, the TE integral for the valence band is
\begin{align}
  \label{eq:crtaval}
  \mathcal{L}_{i,v} =& -\frac{g \tau_0}{4\pi\hbar^2 L}(k_BT)^{i+1}
  \Big[\mathcal{F}_{i+1,v}(\eta+
    \widetilde{\Delta})\nonumber\\
    &+\eta\mathcal{F}_{i,v}(\eta+\widetilde{\Delta})
    -\widetilde{\mathcal{G}}_{i,v}
    (\widetilde{\Delta},\eta)\Big].
\end{align}
The remaining calculation procedure to obtain the TE properties is the
same as in the main text.  Due to the complexity of $\mathcal{L}_i$
and $\mathcal{G}_i$, the integration is performed numerically.  Note
that, in addition to $\mathcal{F}_0$, $\mathcal{F}_1$, and
$\mathcal{F}_2$, we also need $\mathcal{F}_3$ according to the
recursive relation of $\mathcal{L}_i$ in Eqs.~\eqref{eq:crtacon}
and~\eqref{eq:crtaval}.  We obtain
\begin{align}
  \mathcal{F}_{3,c}(\eta) =& \eta^2
  \left[\frac{\eta}{1+e^\eta}+3\ln\left(1+e^{-\eta}\right)\right]\nonumber\\
  &-6\eta\mathrm{Li}_2(-e^{-\eta})-6\mathrm{Li}_3(-e^{-\eta})
\end{align}
and
\begin{align}
  \mathcal{F}_{3,v}(\eta) =& \eta^2
 \left[-\frac{\eta}{1+e^\eta}-3\ln(1+e^{-\eta})\right]\nonumber\\
  &+6\eta\mathrm{Li}_2(-e^{-\eta})+6\mathrm{Li}_3(-e^{-\eta}).
\end{align}

\begin{figure}[t!]
  \centering \includegraphics[clip,width=8cm]{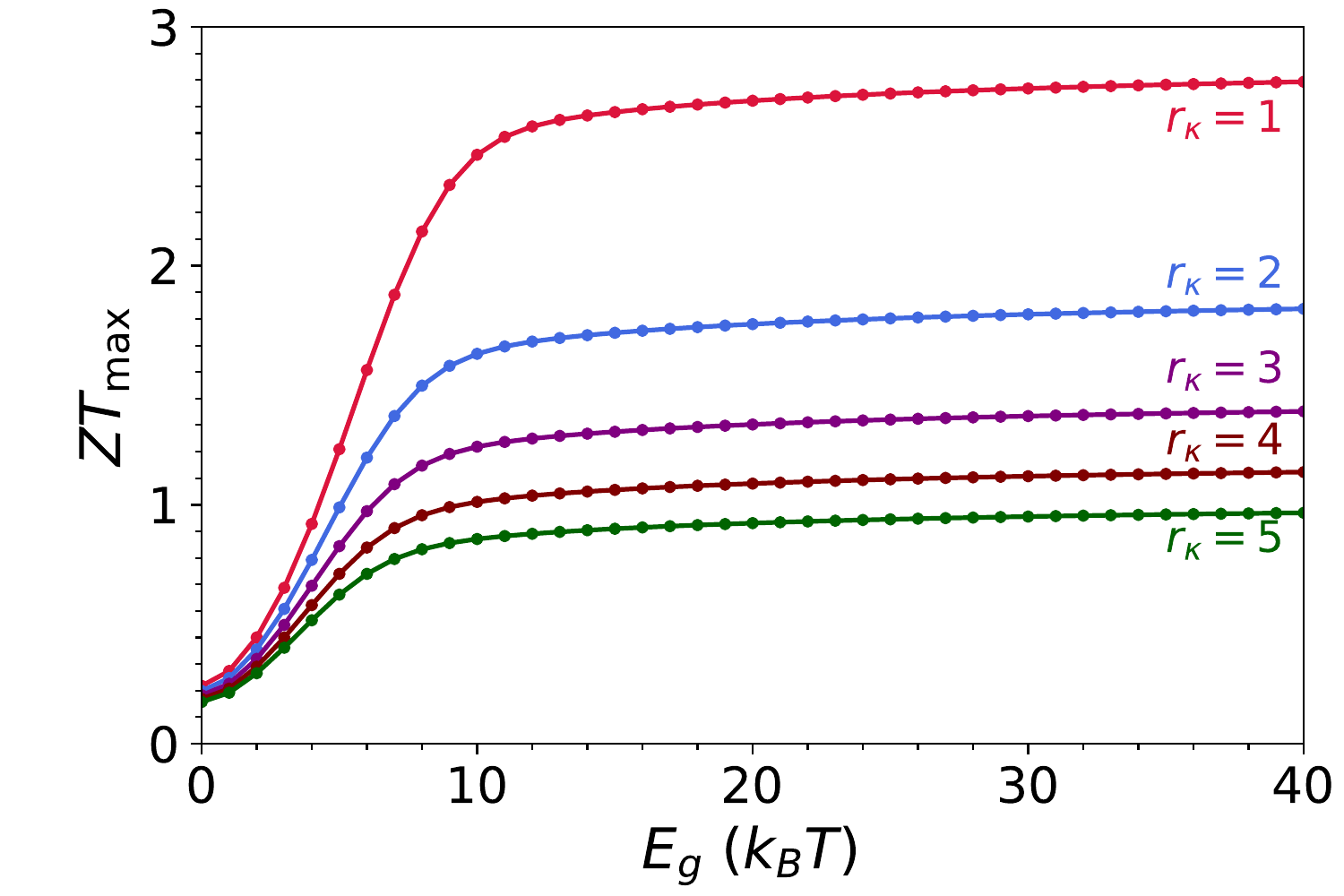}
  \caption{\label{CRTfigZTopt} Maximum $ZT$ values versus band gaps of
    2D Dirac materials within CRTA for several different lattice
    thermal conductivity parameters $r_\kappa$.}
\end{figure}

Figures~\ref{CRTfigTEgapless} and \ref{CRTfigTEgapped} show the
results of TE properties of the 2D Dirac materials within the CRTA by
using the same parameters as in Figs.~\ref{figTEgapless} (gapless,
$\widetilde{\Delta} = 0$) and \ref{figTEgapped} (gapped,
$\widetilde{\Delta} = 1$).  Within the CRTA, we can see that all the
TE properties are overestimated.  The most notable feature is the
nonzero $S$ for the gapless 2D Dirac material
[Fig.~\ref{CRTfigTEgapless}(a)], which leads to the finite
$ZT$~[Fig.~\ref{CRTfigTEgapless}(d)] for different $r_\kappa$ values.
Similarly, $ZT$ for the gapped 2D Dirac material within the CRTA is
also larger than that shown in the main text.  The $ZT_{\textrm{max}}$
profile as a function of band gap for this approximation is shown in
Fig.~\ref{CRTfigZTopt}.  There is no clear peak of
$ZT_{\textrm{max}}$, but the starting gaps to enhance
$ZT_{\textrm{max}}$ are still in the range of $6$--$18k_BT$.

\section{Comparison with first-principles calculation}
\label{app:dft}

The two-band model developed in this work does not only grasp the
essence of electronic contribution to the TE properties of 2D Dirac
materials, but also reasonably fits with calculation results obtained
from the density functional theory (DFT) packages.  We perform band
structure calculation of a relaxed $\rm MoS_2$ structure using
Perdew-Burke-Ernzerhof exchange-correlation functionals~\cite{pbe96},
as implemented in \texttt{Quantum} ESPRESSO, one of the most popular
DFT packages~\cite{QE-2017}.  To achieve the convergence of total
energy calculation, a kinetic-energy cutoff of at least
$\rm E_{cut} = 816$~eV is set for the plane-wave basis set.  The
periodic boundary conditions are employed and a sufficiently large
vacuum layer of $20~\rm\AA$ in the z-direction is adopted so as to
avoid interaction between the adjacent layer.  We obtain the lattice
constant $a=3.19~\rm\AA$ and the band gap $E_g = 1.65$ eV without
spin-orbit coupling.  The TE properties are then calculated by using
the BoltzTraP code~\cite{madsen2006boltztrap}, which is relevant for
comparison with the two-band model within CRTA because the BoltzTrap
code also utilizes a constant value of relaxation time.

\begin{figure}[t]
  \centering \includegraphics[width=8cm]{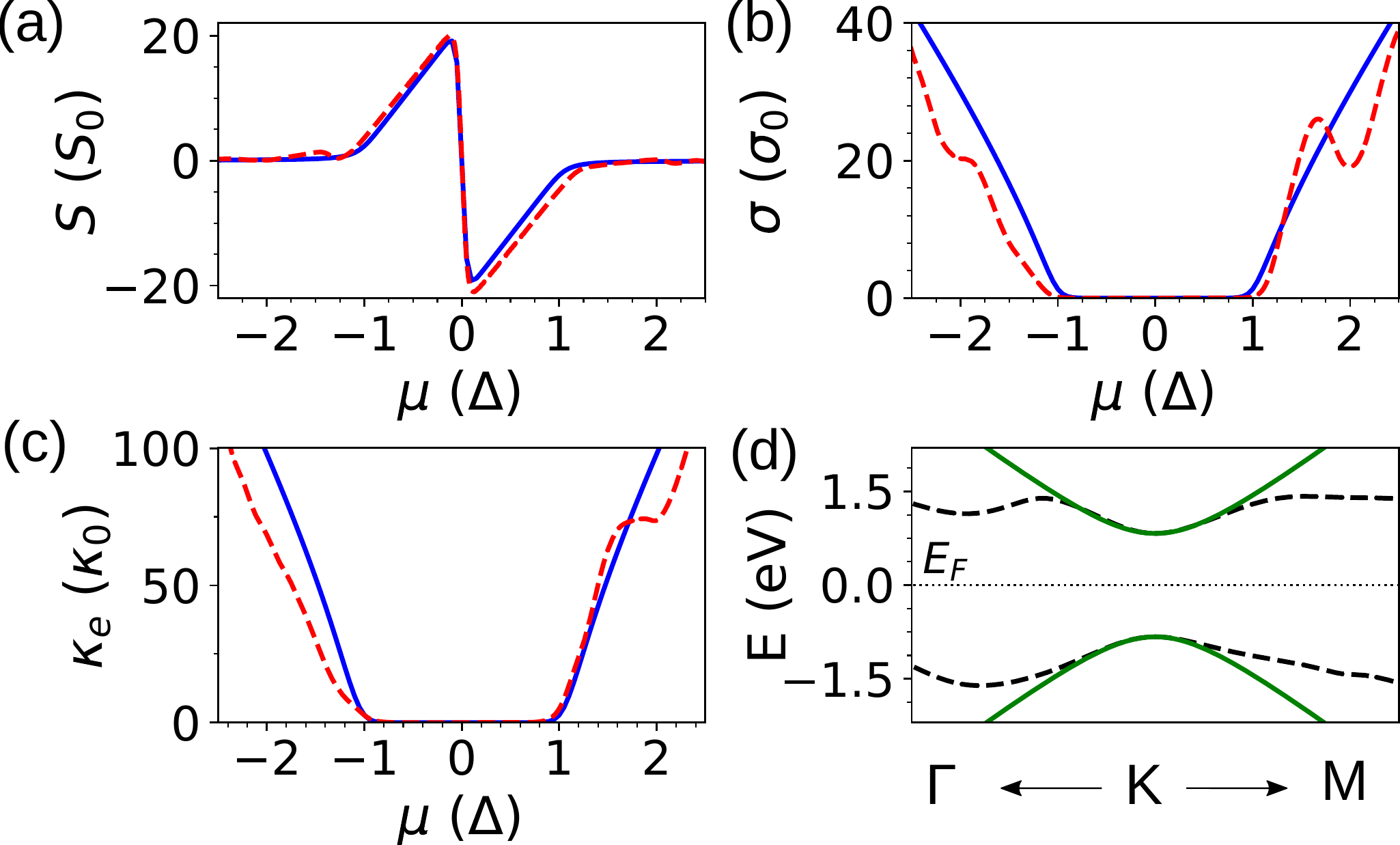}
  \caption{\label{figFit} Thermoelectric properties of $\rm MoS_2$ in
    constant relaxation time approximation. (a) Seebeck coefficient,
    (b) electrical conductivity, and (c) electronic thermal
    conductivity obtained from DFT (dashed lines) are compared with
    our two-band model within CRTA (solid lines).  Here the
    temperature is fixed at $500$ K. (d) $\rm MoS_2$ band structure
    near the K point obtained from DFT calculation (dashed lines) is
    fitted with energy dispersion of gapped 2D Dirac material
    following Eq.~\eqref{eq:dirac} (solid lines) along a certain
    direction of high-symmetry points.  }
\end{figure}

In Figs.~\ref{figFit}(a)--(c) we show TE properties of $\rm MoS_2$
obtained from DFT and from the two-band model within CRTA at $500$~K.
Although the energy dispersion of gapped 2D Dirac materials fits only
in small region near the K point [Fig.~\ref{figFit}(d)], the Seebeck
coefficient, electrical conductivity, and electronic thermal
conductivity from our model fit reasonably well with those from
DFT/BoltzTraP code.  These results indicate that the TE properties,
especially the Seebeck coeficient, of gapped 2D Dirac materials depend
primarily on the size of the gap.  The relatively small discrepancies
between our model and the results of DFT/BoltzTraP can be attributed
simply to the multiband effect and nonlinear energy dispersion far
from the K point.

\section{Comparison with parabolic bands}
\label{app:para}

The curvature of Dirac band is coupled with the size of the gap. In
order to unravel this combined factors to thermoelectricity, we
compare TE of the Dirac bands with that of parabolic bands. The energy
dispersion of parabolic bands takes a form of
$E(k)=\pm \left[\hbar^2|\mathbf{k}|^2/(2m^*)+\Delta\right]$, where
$m^*$ is the effective mass, which accounts the curvature that are
decoupled from the gap $E_g = 2\Delta$.  The 2D parabolic bands have a
constant DOS.  As a result, the CRTA and
$\tau(E)\propto \mathrm{DOS}^{-1}$ approximation are equivalent.

\begin{figure}[t]
  \centering \includegraphics[clip,width=8cm]{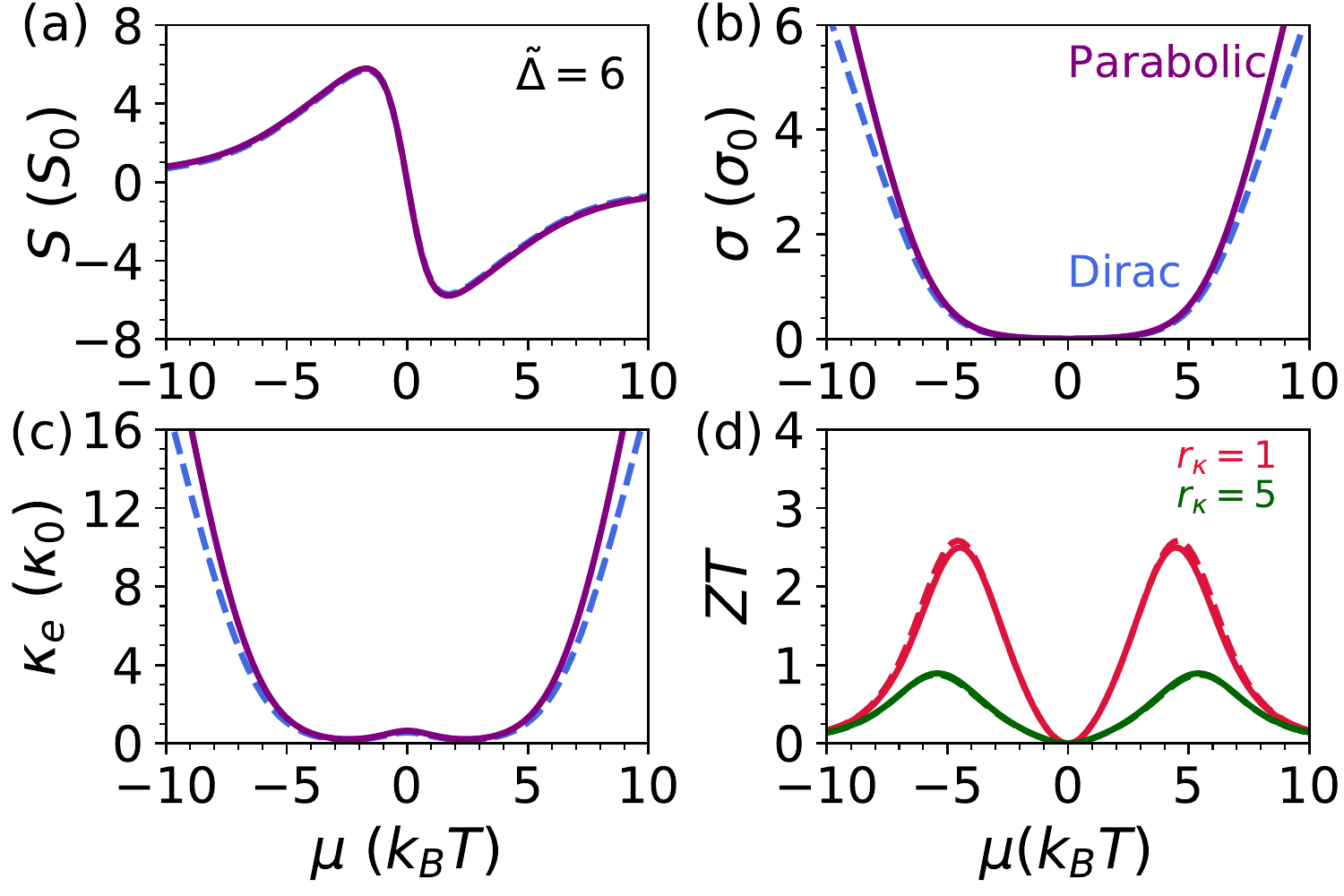}
  \caption{\label{dirac-para} Comparison of thermoelectric properties
    of 2D parabolic bands (solid lines) with Dirac bands (dashed
    lines) within the CRTA for $\widetilde{\Delta} = 6$ (equivalently,
    band gap $E_g = 12 k_B T$).  (a) Seebeck coefficient in units of
    $S_0 = k_B / e$.  (b) Electrical conductivity in units of
    $\sigma_0 = g \tau_0 e^2 k_B T/(4\pi\hbar^2 L)$.  (c) Electron
    thermal conductivity in units of
    $\kappa_0 = g \tau_0 k_B^3 T^2/(4\pi\hbar^2 L)$.  (d)
    Dimensionless figure of merit calculated with two different
    parameters $r_\kappa$ (representing lattice thermal
    conductivity).}
\end{figure}

The TE integral for parabolic bands are:
\begin{align}
  \mathcal{L}_{i,c} =&\frac{g \tau_0}{2\pi\hbar^2 L}(k_BT)^{i+1}
                       \Big[\mathcal{F}_{i+1,c}(\eta-
                       \widetilde{\Delta})\nonumber\\
                     &+(\eta-\widetilde{\Delta})\mathcal{F}_{i,c}
                       (\eta-\widetilde{\Delta})\Big],\label{eq:lparac}\\
  \mathcal{L}_{i,v}= &-\frac{g \tau_0}{2\pi\hbar^2 L}(k_BT)^{i+1}
                       \Big[\mathcal{F}_{i+1,v}(\eta+
                       \widetilde{\Delta})\nonumber\\
                     &+(\eta+\widetilde{\Delta})\mathcal{F}_{i,v}
                       (\eta+\widetilde{\Delta})\Big].\label{eq:lparav}
\end{align}
The remaining procedure to calculate the TE properties is the same as
in the main text.  We again perform numerical calculation for all the
integrals above.  Figures~\ref{dirac-para}(a)-(d) show comparison of
TE quantities for the 2D parabolic-band (solid lines) and Dirac
materials (dashed lines) within the CRTA. The TE quantities are nearly
unchanged for different shape of bands. These results are obvious
because the contribution of relaxation from each energy state is
assumed to be constant in the CRTA.

\begin{figure}[t!]
  \centering \includegraphics[clip,width=8cm]{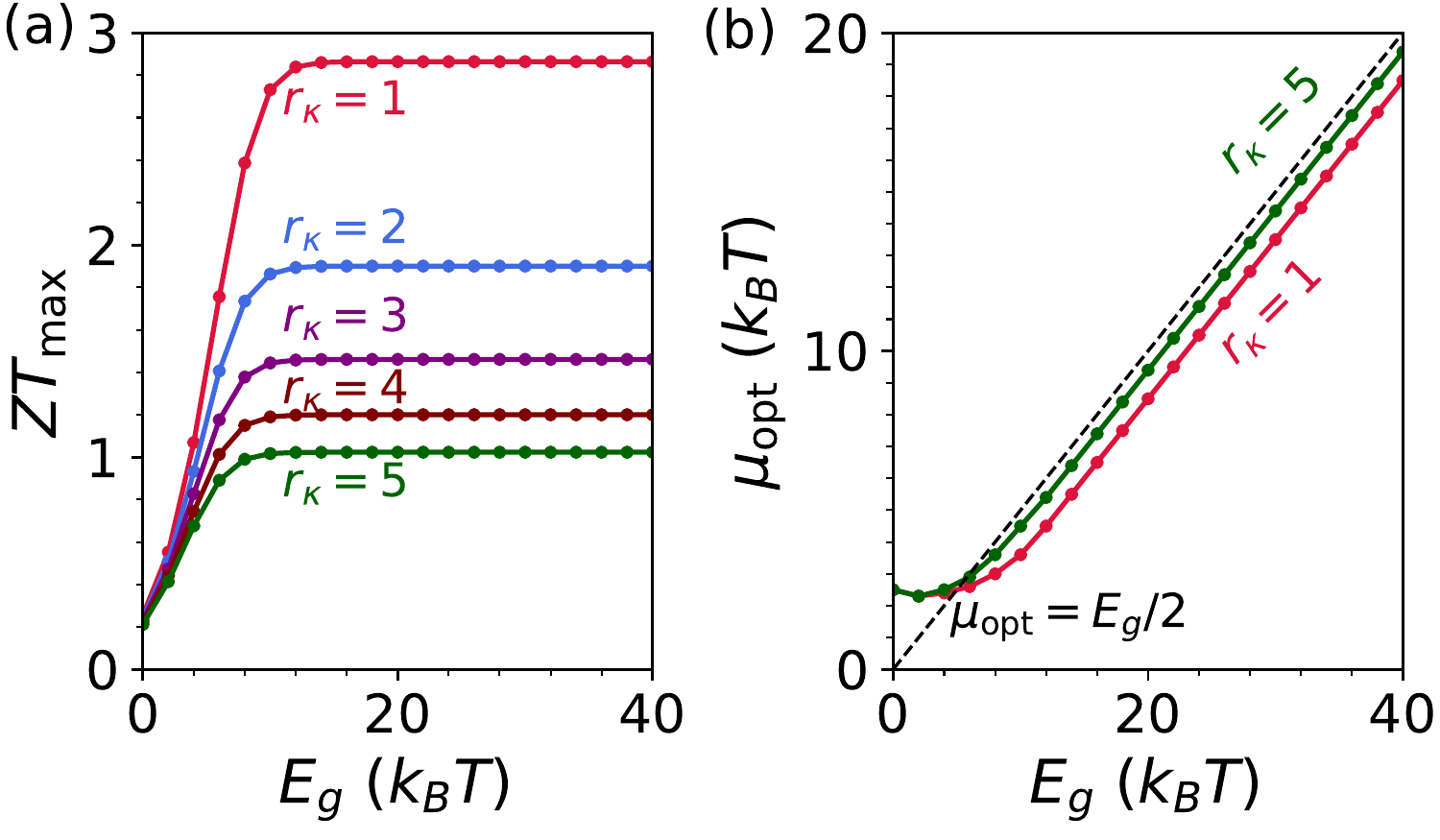}
  \caption{\label{paraZTopt} (a) Maximum $ZT$ values versus band gap
    $E_g$ of parabolic bands for several different lattice thermal
    conductivity parameters $r_\kappa$.  (b) Optimal chemical
    potential $\mu$ that gives maximum $ZT$ as a function of
    $E_g$. The dashed line is a condition when chemical potential is
    located at the band edge ($\mu=E_g/2$).}
\end{figure}

We then plot $ZT_{\rm max}$ vs energy gap ($E_g=2\Delta$) by scanning
through chemical potential $\mu$. The result is shown in
Fig.~\ref{paraZTopt}(a).  $ZT_{\rm max}$ monotonically increases as
$E_g$ increases up to $10k_B T$ and saturates at $E_g$ above
$10k_B T$. This saturation is expected because we assume no gap
dependent relaxation time (mobility).  Figure~\ref{paraZTopt}(b) shows
the optimal chemical potential $\mu$ that gives maximum ZT.  It is
apparent that $\mu_{\rm opt}$ is located very close to the band edge,
indicated by $\mu_{\rm opt}=E_g/2 = \Delta$ as the dashed line. The
flat profile of $ZT_{\rm max}$ vs $E_g$ can be understood from
Eq.~\eqref{eq:lparac} that $\mathcal{L}_{i,c}$ becomes a constant when
$\eta =\widetilde{\Delta}$.  We note that when $\mu$ is at the edge of
conduction or valence band, all systems are doped with the same charge
concentration irrespective of the band gap entering the regime of
degenerate semiconductors.  As a result, $ZT$ is flat at large band
gap.  These results are qualitatively similar to the Dirac band with
CRTA (see Fig.~\ref{CRTfigZTopt}) and to 3D
semiconductors~\cite{sofo94-optgap}.

We, therefore, define the optimal bandgap at the turning point before
$ZT_{\rm max}$ saturates, because increasing the band gap would
require larger $\mu$, which is harder to achieve.  For the 2D
parabolic-band materials, the optimal band gaps with such a definition
are located around 10 $k_B T$.  Furthermore, we should note that
$\tau$ independent of band gap is unrealistic because highly doped
semiconductors are generally suffers large impurity scattering
resulting in the decrease of mobility and $ZT$~\cite{yu-mos2-trans}.
We conclude that the thermoelectric properties are relatively
insensitive to the detailed shape of band structures as long as the
CRTA is concerned.  The change of $\tau$ approximation will affect the
$ZT_{\rm max}$ vs $E_g$ profile for non-parabolic bands (such as the
Dirac band) but it does not change the optimal band gap substantially.


%

\end{document}